\documentclass[12pt]{article}

\usepackage{amsfonts}

\begin{document}

\title{Dark states in quantum photosynthesis}

\author{S.V.Kozyrev, I.V.Volovich \\
Steklov Mathematical Institute}

\maketitle

\begin{abstract}
We discuss a model of quantum photosynthesis with degeneracy in the light-harvesting system. We consider interaction of excitons in chromophores with light and phonons (vibrations of environment). These interactions have dipole form but are different (are related to non--parallel vectors of ''bright'' states). We show that this leads to excitation of non-decaying ''dark'' states. We discuss relation of this model to the known from spectroscopical experiments phenomenon of existence of photonic echo in quantum photosynthesis.
\end{abstract}

\section{Introduction}

Open quantum systems were investigated in many works, in particular, the stochastic limit method was developed \cite{AcLuVo}. Degenerate open systems (when the behavior of quantum systems is more complex) were discussed in particular in \cite{notes}.

In this paper we consider a model of quantum photosynthesis (excitation, transport and absorption of excitons in light-harvesting systems) as a three level quantum system with energy levels $\varepsilon_0$, $\varepsilon_1$, $\varepsilon_2$, $\varepsilon_0<\varepsilon_1<\varepsilon_2$, which interacts with three quantum fields (the reservoirs). The upper level  $\varepsilon_2$ (which corresponds to one exciton states in the system of chromophores) is degenerate. The two lower levels $|0\rangle$, $|1\rangle$ are non degenerate and describe the state ''exciton in the reaction center'' (state $|1\rangle$ with energy $\varepsilon_1$) and the state without excitons $|0\rangle$ with energy $\varepsilon_0$. Three possible transitions between the energy levels are paired to different reservoirs, which describe correspondingly light, phonons (vibrations of protein matrix) and absorption of excitons in the reaction center. Here the light can be coherent.

We consider the dynamics of open quantum systems with dissipation \cite{Hol}, i.e. interaction with any of the reservoirs is described by a generator for density matrices in the Lindblad form, and for the light reservoir we also have an additional term in the generator which describes the interaction with laser. Interaction of degenerate systems with fields generate the so called dark states (or dark--state polaritons) known in quantum optics \cite{Scully}, \cite{dark-state}. Moreover for different reservoirs the corresponding spaces of bright and dark states can be different.

We conjecture that the interactions of chromophores with photons and phonons are different and the bright photonic states generated by interaction with light will contain a dark component for interaction with phonons. States of this kind will not decay in the process of transport of excitons. We discuss the relation of these non-decaying dark states and known from experiments with photonic echo \cite{Engel}, \cite{SFOG} quantum coherencies in photosynthesis systems.

Dynamics in the model under consideration is described by a sum of three generators --- photonic, phononic and the generator of absorption of excitons. We consider the following scheme of computation of manipulation of quantum states of the system.

1) First, we consider the approximation of strong light and weak transport. In this approximation we take into account only the photonic generator and neglect the effects of transport and absorption.

2) Second, we switch off the light and for the obtained at the previous stage photonic stationary state we consider joint action of generators of transport and absorption. After this the upper level of the system will contain only the part of the state which is dark with respect to the transport generator.

3) For the obtained at stage 2 state we consider interaction with a coherent light, which allows to investigate the possibility of photonic echo for this state.

In non degenerate case (when energy level $\varepsilon_2$ is non degenerate) the considered in this paper non-equilibrium quantum system was discussed in \cite{noneqlaser}. The flow of excitons in the system was computed and excitation of quantum coherencies for interaction with laser was considered. When the laser is switched off these coherencies (in non degenerate case) decay due to interaction with phonons. Quantum dynamics in the stochastic limit in presence of laser field was studied in \cite{lambdaatom}. Properties of antibunched light are discussed for instance in \cite{Vol}.

In paper \cite{tmf2014} effects of degeneracy in exciton transport were investigated by the stochastic limit method. It was shown that, using constructive interference, it is possible to achieve quantum amplification of exciton transport (the supertransport effect). The possibility of excitation of non-decaying dark states in a degenerate system can be considered as a side effect of the supertransport.

In \cite{Novoderezhkin} dark states in photosynthetic systems were studied experimentally.

In \cite{OhyaVolovich} application of quantum methods to computations and biology was discussed. In \cite{DongXu} dark states in photosynthesis were considered but dark states in this paper were considered in a different way compared to the present paper (as collective states of excitons, photons and phonons), in particular interactions between excitons and photons, excitons an phonons in \cite{DongXu} are similar (in this case the effect considered in the present paper does not take place).

The exposition of the present paper is as follows. In section 2 the Hamiltonian of light-harvesting system interacting with three reservoirs (photons, phonons and absorption) is described. In section 3 the corresponding generators of dynamics of density matrix of the system are considered (there are three such generators which describe excitation, transport and absorption of excitons). In section 4 bright, dark and off-diagonal states for the mentioned generators are discussed. In section 5 preparation of quantum states by interaction with light (including laser field) is considered. In section 6 the state obtained in section 5 is subjected to transport and absorption (when the light is switched off). In section 7 we consider the interaction of the obtained at the previous stage dark state of the system with laser field and discuss the relation to photonic echo experiments.

\section{Hamiltonian of the model}

We consider a system with three energy levels $\varepsilon_0<\varepsilon_1<\varepsilon_2$, where the upper level is degenerate, with the Hamiltonian
\begin{equation}\label{H_S}
H_S=\varepsilon_0 |0\rangle\langle 0|+ \varepsilon_1 |1\rangle\langle 1|+\varepsilon_2\sum_{j=2}^{N} |j\rangle\langle j|.
\end{equation}

This Hamiltonian describes a light-harvesting system, $|0\rangle$ corresponds to a state without excitons, $|1\rangle$ is a state ''exciton in the reaction center'', $|j\rangle$ correspond to one-exciton states of chromophores.

System interacts with three quantum fields (reservoirs) in a dipole way. Transitions between the levels with energies $\varepsilon_0$ and $\varepsilon_2$ (in particular creation of excitons) are related to interaction with light (Bose field in the Gibbs state with the temperature $\beta_{\rm em}^{-1}=6000K$, or laser field with the frequency $\varepsilon_2-\varepsilon_0$), transitions between the levels $\varepsilon_2$ and $\varepsilon_1$ (transport of excitons to the reaction center) are related to interaction with phonons, or vibrations of protein matrix (described by the Bose field with the temperature $\beta_{\rm ph}^{-1}=300K$), and transitions between the levels $\varepsilon_1$ and $\varepsilon_0$ (absorption of excitons in the reaction center) are described by interaction with the sink reservoir in the Fock state (i.e. reservoir with the zero temperature).

Thus we have three reservoirs described by Hamiltonians of quantum Bose fields $H_{\rm em}$ (light, or electromagnetic field), $H_{\rm ph}$ (phonons, or vibrations of protein matrix), $H_{\rm sink}$ (sink, or absorption of excitons in the reaction center), each of the reservoir Hamiltonians has the form
$$
H_R=\int_{\mathbb{R}^3} \omega_R(k)a^{*}_R(k) a_R(k) dk,
$$
where $R={\rm em},\, {\rm ph},\, {\rm sink}$ enumerate the reservoirs, $\omega_R$ is the dispersion of the Bose field $a_R$.

Each of the reservoirs is in a mean zero Gaussian state with the quadratic correlator
$$
\langle a^{*}_{R}(k)a_R(k') \rangle=N_R(k)\delta(k-k').
$$

Here $N_R(k)$ (number of the field quanta with wave number $k$) is equal to (for the inverse temperature $\beta_R$)
\begin{equation}\label{temperature}
N_R(k)={1\over{e^{\beta_R\omega_R(k)}-1}}.
\end{equation}

The full Hamiltonian has the form
$$
H=H_S+H_{\rm em}+H_{\rm ph}+H_{\rm sink}+ \lambda\left( H_{I,{\rm em}}+H_{I,{\rm ph}}+H_{I,{\rm sink}}\right),
$$
where the interaction Hamiltonians $H_{I,{\rm em}}$, $H_{I,{\rm ph}}$, $H_{I,{\rm sink}}$ have dipole forms and are given by formulae (\ref{H_em}), (\ref{H_ph}), (\ref{H_sink}) below.

Interaction of the system with light is described by the Hamiltonian
\begin{equation}\label{H_em}
H_{I,{\rm em}} = A_{\rm em}|\chi\rangle\langle 0| + A^{*}_{\rm em}|0\rangle\langle \chi|,\qquad A^{*}_{\rm em}=\int_{\mathbb{R}^3} g_{\rm em}(k)a^{*}_{\rm em}(k)  dk,
\end{equation}
where the bright photonic state $\chi$ belongs to the level with energy $\varepsilon_2$, this state can be taken normalized $\|\chi\|=1$, $g_{\rm em}(k)$ is the form--factor of the field.

Transport of excitons to the reaction center is related to interaction of the system with phonons
\begin{equation}\label{H_ph}
H_{I,{\rm ph}} = A_{\rm ph}|\psi\rangle\langle 1| + A^{*}_{\rm ph}|1\rangle\langle \psi|,\qquad A^{*}_{\rm ph}=\int_{\mathbb{R}^3} g_{\rm ph}(k)a^{*}_{\rm ph}(k)  dk,
\end{equation}
here the bright phononic state $\psi$ with the energy $\varepsilon_2$ is normalized $\|\psi\|=1$.

Vectors $\psi$ and $\chi$ belong to the same degenerate level with energy $\varepsilon_2$ (corresponding to excitons on chromophores).
Crucial feature of the model under consideration is as follows --- vectors $\psi$ and $\chi$ are non-parallel.

Absorption of excitons in the reaction center is described by interaction with the additional field of sink (with the zero temperature)
\begin{equation}\label{H_sink}
H_{I,{\rm sink}} = A_{\rm sink}|1\rangle\langle 0| + A^{*}_{\rm sink}|0\rangle\langle 1|,\qquad A^{*}_{\rm sink}=\int_{\mathbb{R}^3} g_{\rm sink}(k)a^{*}_{\rm sink}(k)  dk.
\end{equation}

\medskip

\noindent{\bf Remark.}\quad Usually in papers on photosynthesis some different Hamiltonian is used. In particular in this Hamiltonian the part (\ref{H_S}) is non-diagonal and contains terms corresponding to the dipole interaction of excitons, the interaction Hamiltonian is also different. Here we consider the result of diagonalization of the system Hamiltonian $H_S$ (transition to the so called ''global'' basis), i.e. in our notations the states  $|j\rangle$of the system correspond to the states from the ''global'' basis. Relation between the ''global'' and ''local'' approaches in theory of open quantum systems was discussed in \cite{TruVol}.

\section{Generators of the dynamics}

For investigation of the model we will use the method of quantum stochastic limit \cite{AcLuVo}, \cite{notes}. In this limit dynamics of reduced density matrix of a system interacting with reservoir is generated by operator in the Lindblad form, see formulae (\ref{theta_em}), (\ref{theta_ph}), (\ref{theta_sink}) below. Quantum dynamics in the stochastic limit in presence of a laser field was discussed in \cite{lambdaatom}.

For the considered model with three reservoirs the dynamics will be generated by a sum of three generators
$$
\frac{d}{dt}\rho(t)=\left(\theta_{\rm em}+i[\cdot, H_{\rm eff}]+\theta_{\rm ph}+\theta_{\rm sink}\right)(\rho(t)).
$$

We will manipulate quantum states of the system by switching different generators in this formula. Actually it is possible to manipulate only with the photonic generator $\theta_{\rm em}+i[\cdot, H_{\rm eff}]$.

Creation and annihilation of excitons are described by the photonic generator equal to a sum of the dissipative Lindblad term and the term related to coherent field \cite{AcLuVo, notes}:
\begin{equation}\label{L+theta_em}
L_{\rm em}=\theta_{\rm em}+i[\cdot, H_{\rm eff}],\quad H_{\rm eff}=s(|\chi\rangle\langle 0|+|0\rangle\langle \chi|),\qquad s\in\mathbb{R}.
\end{equation}
\begin{equation}\label{theta_em}
\theta_{\rm em}(\rho)=
2\gamma^{-}_{{\rm re,em}}
\left(
\langle \chi|\rho|\chi\rangle |0\rangle\langle 0|
-{1\over 2}
\{\rho,|\chi\rangle\langle \chi|\}\right)
-i\gamma^{-}_{{\rm im,em}} [\rho,|\chi\rangle\langle \chi|]+
$$
$$
+
2\gamma^{+}_{{\rm re,em}}
\left(\langle 0| \rho |0\rangle |\chi\rangle\langle \chi|
-{1\over 2}
\{\rho,|0\rangle\langle 0| \}\right)
+i\gamma^{+}_{{\rm im,em}} [\rho,|0\rangle\langle 0|].
\end{equation}

The constants $\gamma$ have the form
\begin{equation}\label{Regg+1}
\gamma^{+}_{\rm re, em}=\pi\int
|g_{\rm em}(k)|^2\delta(\omega_{\rm em}(k)-\varepsilon_2+\varepsilon_0)N_{\rm em}(k)dk,
\end{equation}
\begin{equation}\label{Regg-1}
\gamma^{-}_{\rm re, em}=\pi\int
|g_{\rm em}(k)|^2\delta(\omega_{\rm em}(k)-\varepsilon_2+\varepsilon_0)(N_{\rm em}(k)+1)dk,
\end{equation}
\begin{equation}\label{Imgg+1}
\gamma^{+}_{\rm im, em}=-\int
|g_{\rm em}(k)|^2\,{\rm P.P.}\,{1\over\omega_{\rm em}(k)-\varepsilon_2+\varepsilon_0}N_{\rm em}(k)dk,
\end{equation}
\begin{equation}\label{Imgg-1}
\gamma^{-}_{\rm im, em}=-\int
|g_{\rm em}(k)|^2\,{\rm P.P.}\,{1\over\omega_{\rm em}(k)-\varepsilon_2+\varepsilon_0}(N_{\rm em}(k)+1)dk.
\end{equation}
Here the function $N_{\rm em}(k)$ is given by (\ref{temperature}), $\beta_{\rm em}^{-1}=6000K$ for the sun light.

For the laser field $N_{\rm em}(k)=0$, this implies $\gamma^{+}_{\rm re, em}=\gamma^{+}_{\rm im, em}=0$, but $\gamma^{-}_{\rm re, em},\gamma^{-}_{\rm im, em}\ne 0$ (i.e. dissipative part of the generator is non zero even for coherent field).

Transport of excitons is described by the phononic generator, here $\beta_{\rm ph}^{-1}=300K$
\begin{equation}\label{theta_ph}
\theta_{\rm ph}(\rho)=
2\gamma^{-}_{{\rm re,ph}}
\left(
\langle \psi|\rho|\psi\rangle |1\rangle\langle 1|
-{1\over 2}
\{\rho,|\psi\rangle\langle \psi|\}\right)
-i\gamma^{-}_{{\rm im,ph}} [\rho,|\psi\rangle\langle \psi|]+
$$
$$
+
2\gamma^{+}_{{\rm re,ph}}
\left(\langle 1| \rho |1\rangle |\psi\rangle\langle \psi|
-{1\over 2}
\{\rho,|1\rangle\langle 1| \}\right)
+i\gamma^{+}_{{\rm im,ph}} [\rho,|1\rangle\langle 1|].
\end{equation}

Absorption of excitons is described by the sink generator with $\beta_{\rm sink}^{-1}=0K$
\begin{equation}\label{theta_sink}
\theta_{\rm sink}(\rho)=
2\gamma^{-}_{{\rm re,sink}}
\left(
\langle 1|\rho|1\rangle |0\rangle\langle 0|
-{1\over 2}
\{\rho,|1\rangle\langle 1|\}\right)
-i\gamma^{-}_{{\rm im,sink}} [\rho,|1\rangle\langle 1|].
\end{equation}
Here coefficients $\gamma^{+}$ are equal to zero since the temperature is zero.

Constants $\gamma$ in (\ref{theta_ph}), (\ref{theta_sink}) are given by formulae analogous to (\ref{Regg+1}), (\ref{Regg-1}), (\ref{Imgg+1}), (\ref{Imgg-1}) with different Bohr frequencies (differences of energies of the levels), dispersions and temperatures of the reservoirs (i.e. we substitute these parameters by the corresponding for phonons and sink).

\section{Bright, dark and off-diagonal states}

For Lindblad generators of the form (\ref{theta_em}), (\ref{theta_ph}), (\ref{theta_sink}) one can consider expansion of the space of states of the system in a sum of orthogonal subspaces of bright, dark and off-diagonal states. Bright and dark states were extensively discussed in quantum optics, see \cite{Scully}, \cite{dark-state}. We use here the approach and notations of \cite{tmf2014}.

These subspaces depend on the generator and are different for different generators. Let us discuss the photonic generator (\ref{theta_em}).
Generator $\theta_{\rm em}$ describes creation and annihilation of excitons by interaction of chromophores and electromagnetic field. Bright mixed states for this generators are linear combinations of matrices
$$
|0\rangle\langle 0|,\quad |\chi\rangle\langle \chi|.
$$
Below we usually will ignore the positivity and normalization for states, i.e. we will talk about matrices instead of states (even if we will use the term state).

Pure bright state of the upper level (with energy $\varepsilon_2$) is the vector $|\chi\rangle$ in (\ref{H_em}). Pure dark states of the upper level are vectors $|\phi\rangle$ orthogonal to the pure bright state: $\phi\bot\chi$.

Mixed dark states are matrices $B$ which give zero when multiplied by any mixed bright state $A$, i.e.
$$
AB=BA=0.
$$

In the case of the generator $\theta_{\rm em}$ these states are linear combinations of matrices
$$
|\phi\rangle\langle \phi'|,\quad |1\rangle\langle 1|,\quad |\phi\rangle\langle 1|,\quad |1\rangle\langle \phi|,\quad \phi\bot\chi, \phi'\bot\chi.
$$

Off-diagonal matrices are matrices $C$ orthogonal to all bright matrices $A$ and all dark matrices $B$ with respect to the scalar product $(\cdot,\cdot)={\rm tr}(\cdot\cdot)$, i.e.
$$
{\rm tr}(CA)={\rm tr}(CB)=0.
$$

This subspace contains matrices corresponding to transitions between bright and dark subspaces, and between levels in the bright subspace, for the generator $\theta_{\rm em}$ the off-diagonal subspace is generated by matrices
$$
|\chi\rangle\langle 0|,\quad |\chi\rangle\langle \phi|,\quad |\chi\rangle\langle 1|,\quad |1\rangle\langle 0|,\quad |\phi\rangle\langle 0|,\quad \phi\bot\chi
$$
and conjugated.

Dark states described above are stationary with respect to the dynamics generated by $\theta_{\rm em}$. The bright space corresponds to processes of creation and annihilation of excitons (for the generator $\theta_{\rm em}$) and to process of transport of excitons (for the generator $\theta_{\rm ph}$). Off-diagonal matrices decay which corresponds to the decoherence phenomenon. In presence of a coherent field (generator $i[\cdot, H_{\rm eff}]$) off-diagonal matrices can be created.

Described expansion of the state of matrices depends on the generator and for the phononic generator $\theta_{\rm ph}$ will be completely different. In particular matrix $|1\rangle\langle 1|$ is dark for $\theta_{\rm em}$ and bright for $\theta_{\rm ph}$, and matrix $|\chi\rangle\langle \chi|$ is bright for $\theta_{\rm em}$ and contains a combination of bright, dark and off-diagonal terms for $\theta_{\rm ph}$ (since the phononic pure bright vector $\psi$ is not parallel to the photonic pure bright vector $\chi$).

\medskip

Let us discuss also bright, dark and off-diagonal matrices for the phononic generator $\theta_{\rm ph}$ given by (\ref{theta_ph}). The pure bright vector of the upper level is given by $|\psi\rangle$ in (\ref{H_ph}).
Mixed bright matrices for $\theta_{\rm ph}$ are linear combinations of
$$
|1\rangle\langle 1|,\quad |\psi\rangle\langle \psi|.
$$

The space of mixed dark states is generated by matrices
$$
|\eta\rangle\langle \eta'|,\quad |0\rangle\langle 0|,\quad |\eta\rangle\langle 0|,\quad |0\rangle\langle \eta|,\quad \eta\bot\psi, \eta'\bot\psi.
$$

The off-diagonal space is a linear span of matrices
$$
|\psi\rangle\langle 1|,\quad |\psi\rangle\langle \eta|,\quad |\psi\rangle\langle 0|,\quad |0\rangle\langle 1|,\quad |\eta\rangle\langle 1|,\quad \eta\bot\psi
$$
and conjugated.

\section{Stage 1: light is switched on}

Let us discuss the process of manipulation of quantum states of light-harvesting system. The initial state is the density matrix without excitons
$$
\rho_0=|0\rangle\langle 0|.
$$

We apply to this initial state the dynamics generated by the photonic generator $L_{\rm em}=\theta_{\rm em}+i[\cdot, H_{\rm eff}]$ of the form (\ref{L+theta_em}) where the light contains also coherent component. This dynamics  puts the system in a stationary state of the form
\begin{equation}\label{stage1}
\rho_1=\rho_{00}|0\rangle\langle 0|+\rho_{\chi\chi}|\chi\rangle\langle \chi|+\rho_{\chi 0}|\chi\rangle\langle 0|+\rho_{0\chi}|0\rangle\langle \chi|,
\end{equation}
where
\begin{equation}\label{rho00rhochichi}
\rho_{00}=
\frac{\gamma^{-}_{{\rm re,em}}-\frac{s^2}{2}\left(\frac{1}{\mu_{20}}+\frac{1}{\mu_{02}}\right)}
{\gamma^{+}_{{\rm re,em}}+\gamma^{-}_{{\rm re,em}}-s^2\left(\frac{1}{\mu_{20}}+\frac{1}{\mu_{02}}\right)},\qquad
\rho_{\chi\chi}=
\frac{\gamma^{+}_{{\rm re,em}}-\frac{s^2}{2}\left(\frac{1}{\mu_{20}}+\frac{1}{\mu_{02}}\right)}
{\gamma^{+}_{{\rm re,em}}+\gamma^{-}_{{\rm re,em}}-s^2\left(\frac{1}{\mu_{20}}+\frac{1}{\mu_{02}}\right)},
\end{equation}
\begin{equation}\label{rhochi0rho0chi}
\rho_{\chi 0}=\frac{is}{\mu_{20}}\frac{\gamma^{-}_{{\rm re,em}}-\gamma^{+}_{{\rm re,em}}}
{\gamma^{+}_{{\rm re,em}}+\gamma^{-}_{{\rm re,em}}-s^2\left(\frac{1}{\mu_{20}}+\frac{1}{\mu_{02}}\right)},\qquad
\rho_{0\chi}=-\frac{is}{\mu_{02}}\frac{\gamma^{-}_{{\rm re,em}}-\gamma^{+}_{{\rm re,em}} }
{\gamma^{+}_{{\rm re,em}}+\gamma^{-}_{{\rm re,em}}-s^2\left(\frac{1}{\mu_{20}}+\frac{1}{\mu_{02}}\right)},
\end{equation}
\begin{equation}\label{mu20}
\mu_{20}=-\gamma^{-}_{{\rm re,em}}-\gamma^{+}_{{\rm re,em}}+i\gamma^{-}_{{\rm im,em}}+i\gamma^{+}_{{\rm im,em}},
\end{equation}
\begin{equation}\label{mu02}
\mu_{02}=-\gamma^{-}_{{\rm re,em}}-\gamma^{+}_{{\rm re,em}}-i\gamma^{-}_{{\rm im,em}}-i\gamma^{+}_{{\rm im,em}}.
\end{equation}

One can say that we used the approximation of strong light and ignore the contributions to the dynamics from the phononic generator $\theta_{\rm ph}$ and the sink generator $\theta_{\rm sink}$.

In particular, if there is no coherent field then off-diagonal elements of the matrix $\rho_1$ vanish $\rho_{\chi 0}=\rho_{0\chi}=0$ and diagonal elements $\rho_{00}$, $\rho_{\chi\chi}$ are given by the Gibbs state satisfying
$$
\frac{\rho_{\chi\chi}}{\rho_{00}}=e^{-\beta_{\rm em}\left(\varepsilon_2-\varepsilon_0\right)}.
$$

In the limit of strong coherent field $s\to\infty$ we get for (\ref{rho00rhochichi}), (\ref{rhochi0rho0chi})
$$
\rho_{00}=\rho_{\chi\chi}=\frac{1}{2},\qquad \rho_{\chi 0}=\rho_{0\chi}=0.
$$

For the photonic generator $L_{\rm em}=\theta_{\rm em}+i[\cdot, H_{\rm eff}]$ there exist also dark stationary states which are linear combinations of matrices
$$
|\phi\rangle\langle\phi'|,\quad |\phi\rangle\langle 1|,\quad |1\rangle\langle\phi|,\quad |1\rangle\langle 1|,\qquad \phi,\phi'\bot\chi.
$$
Thus addition of a coherent field does not change the space of dark states.

\medskip

\noindent{\bf Proof.}\quad Dynamics with the generator $L_{\rm em}$ and the initial sate $\rho_0=|0\rangle\langle 0|$ belongs to the space of matrices of a form (\ref{stage1}).
In the space of matrices (\ref{stage1}) the matrix of the generator $L_{\rm em}$ is as follows
$$
L_{\rm em}\pmatrix{\rho_{\chi\chi}\cr \rho_{00}\cr \rho_{\chi0}\cr \rho_{0\chi}\cr}=
\pmatrix{
-2\gamma^{-}_{{\rm re,em}};&  2\gamma^{+}_{{\rm re,em}};& is;   &       -is     \cr
2\gamma^{-}_{{\rm re,em}}; & -2\gamma^{+}_{{\rm re,em}};& -is;  &        is     \cr
                     is;&                     -is;& \mu_{20};&      0     \cr
                    -is;&                      is;&     0;&  \mu_{02}     \cr
}
\pmatrix{\rho_{\chi\chi}\cr \rho_{00}\cr \rho_{\chi0}\cr \rho_{0\chi}\cr}.
$$

For the stationary state the third equation of the system $L_{\rm em}\rho=0$ implies
$$
is(\rho_{\chi\chi}-\rho_{00})=-\mu_{20}\rho_{\chi 0},
$$
i.e.
$$
\rho_{\chi 0}=-\frac{is}{\mu_{20}}(\rho_{\chi\chi}-\rho_{00}),\qquad \rho_{0\chi}=\frac{is}{\mu_{02}}(\rho_{\chi\chi}-\rho_{00}).
$$

Substitution of these $\rho_{\chi 0}$, $\rho_{0\chi}$ in the first equation of  $L_{\rm em}\rho=0$ gives
$$
\left(2\gamma^{+}_{{\rm re,em}}-s^2\left(\frac{1}{\mu_{20}}+\frac{1}{\mu_{02}}\right)\right)\rho_{00}=
\left(2\gamma^{-}_{{\rm re,em}}-s^2\left(\frac{1}{\mu_{20}}+\frac{1}{\mu_{02}}\right)\right)\rho_{\chi\chi}.
$$
The normalization condition $\rho_{00}+\rho_{\chi\chi}=1$ implies (\ref{rho00rhochichi}), (\ref{rhochi0rho0chi}).

\medskip

\noindent{\bf Remark.}\quad It might seem that the described stationary state does not depend on the degeneracy, in particular the off-diagonal element (\ref{rhochi0rho0chi}) is similar for both cases with and without degeneracy. This formula follows from the normalization $\|\chi\|=1$ in (\ref{H_em}) which is always possible: if $\|\chi\|\ne 1$ then one can the corresponding normalization add to the expression for the field in (\ref{H_em}):
$$
H_{I,{\rm em}} = A_{\rm em}|\chi\rangle\langle 0| + A^{*}_{\rm em}|0\rangle\langle \chi|=\|\chi\| A_{\rm em}\frac{1}{\|\chi\|}|\chi\rangle\langle 0| + \|\chi\| A^{*}_{\rm em}\frac{1}{\|\chi\|}|0\rangle\langle \chi|.
$$

Therefore the normalization $\|\chi\|=1$ means that the parameter $s$ is multiplied by $\|\chi\|$ and values $\gamma^{\pm}_{{\rm em}}$ are multiplied by $\|\chi\|^2$. This transformation gives
$$
\rho_{\chi 0}=\frac{is}{\|\chi\|\mu_{20}}\frac{\gamma^{-}_{{\rm re,em}}-\gamma^{+}_{{\rm re,em}}}
{\gamma^{+}_{{\rm re,em}}+\gamma^{-}_{{\rm re,em}}-\frac{s^2}{\|\chi\|^2}\left(\frac{1}{\mu_{20}}+\frac{1}{\mu_{02}}\right)}.
$$

\section{Stage 2: light is switched off}

The second stage of manipulation of quantum states of light-harvesting system is as follows: we switch off the light, i.e. we use the obtained at the previous step state $\rho_1$ given by (\ref{stage1}), (\ref{rho00rhochichi}), (\ref{rhochi0rho0chi}) as the initial state for dynamics with the generator $\theta_{\rm ph}+\theta_{\rm sink}$ (sum of phononic generator and sink generator).

Let us consider the expansion of the bright photonic state
$$
\chi=\chi_0+\chi_1,\qquad \chi_0 \| \psi,\quad \chi_1 \bot \psi
$$
in the sum of contributions parallel and orthogonal to the bright phononic state $\psi$
$$
|\chi_0\rangle=\langle\psi,\chi\rangle |\psi\rangle =|\psi\rangle\langle\psi| |\chi\rangle,\qquad |\chi_1\rangle=(1-|\psi\rangle\langle\psi|) |\chi\rangle.
$$

Let us substitute this expansion in expression (\ref{stage1}), (\ref{rho00rhochichi}), (\ref{rhochi0rho0chi}) for the stationary photonic state and apply the generator $\theta_{\rm ph}+\theta_{\rm sink}$. Discussion of section 4 of dynamics in spaces of bright, dark and off-diagonal matrices implies that all terms in the expansion which contain $\chi_0$ will decay since the transport generator $\theta_{\rm ph}$ will transfer excitons to the reaction center where excitons will be absorbed. Hence the system will tend to the stationary state of the form
\begin{equation}\label{stage2}
\rho_2=\rho_{00}|0\rangle\langle 0|+\rho_{\chi\chi}|\chi_1\rangle\langle \chi_1|+\rho_{\chi 0}|\chi_1\rangle\langle 0|+\rho_{0\chi}|0\rangle\langle \chi_1|,
\end{equation}
where $\rho_{\chi\chi}$, $\rho_{\chi 0}$, $\rho_{0\chi}$ are given by (\ref{rho00rhochichi}), (\ref{rhochi0rho0chi}), and $\rho_{00}$ is given by the condition that trace of density matrix is equal to one
$$
\rho_{00}=1-\|\chi_1\|^2\rho_{\chi\chi}.
$$

In the model under consideration the obtained state (\ref{stage2}) can exist infinite time (in reality the lifetime should be finite but long). If the bright states for photons and phonons coincide  (i.e. $|\chi_1\rangle=0$) then we will get $\rho_2=|0\rangle\langle 0|$.
In general case of non-parallel $\psi$, $\chi$ the obtained state $\rho_2$ will contain non-decaying dark part. This part can be observed in spectroscopical experiments, see the next section.

\medskip

\noindent{\bf Remark.}\quad Vectors $\psi$, $\chi$ are norm one vectors in multidimensional space. Random independent normed vectors in a multidimensional space are close to orthogonal (since measure of a unit sphere is concentrated near the equator of this sphere).

On the other hand, biological systems undergo selection pressure, for the present model this is the selection pressure to increase effectiveness of exciton transport. This selection pressure will aim to make vectors $\chi$ and $\psi$ parallel (in this case the transport will be most effective). Therefore the angle between $\chi$ and $\psi$ will be related to competition between entropy and selection.

\section{Stage 3: light is switched on}

At the third stage of manipulation of quantum states of excitons we subject the obtained at the previous step state $\rho_2$ of the form (\ref{stage2}) to interaction with coherent light. We ignore transport and absorption (for example the light strong and we consider small times). We will show that in this situation it is possible to observe photonic echo.

Spectroscopy is the application of dynamics with the generator $i[\cdot, H_{\rm eff}]$ to the off-diagonal part of density matrix proportional to $|\chi\rangle\langle 0|-|0\rangle\langle \chi|$, see below. Thus we are interested in the term in (\ref{stage2}) of the form
\begin{equation}\label{offdiag}
\rho_{\chi 0}|\chi_1\rangle\langle 0|+\rho_{0\chi}|0\rangle\langle \chi_1|,
\end{equation}
where the matrix elements $\rho_{\chi 0}$, $\rho_{0\chi}$ are given by (\ref{rhochi0rho0chi}).
In particular this term can be non-zero only if $s\ne 0$, i.e. the state (\ref{stage1}) should be excited by light with coherent component.

Let us recall that $\chi_1$ is the orthogonal complement to projection of $\chi$ (bright photonic vector) to $\psi$ (bright phononic vector). We define $\chi_2$ as projection of $\chi_1$ to $\chi$ (i.e. $\chi_2$ is parallel to $\chi$), then
$$
|\chi_2\rangle=\left(1-|\langle\psi,\chi\rangle|^2\right)|\chi\rangle.
$$

Applying $i[\cdot, H_{\rm eff}]$ to (\ref{offdiag}) we will get non-zero contribution only for the following matrix (since the contribution from orthogonal complement to $|\chi_2\rangle$ vanish)
$$
\rho_{\chi 0}|\chi_2\rangle\langle 0|+\rho_{0\chi}|0\rangle\langle \chi_2|=
$$
$$
={\rm Re}\, \rho_{\chi 0}\left(1-|\langle\psi,\chi\rangle|^2\right) \left(|\chi\rangle\langle 0|+|0\rangle\langle \chi|\right)+
i{\rm Im}\, \rho_{\chi 0}\left(1-|\langle\psi,\chi\rangle|^2\right)\left(|\chi\rangle\langle 0|-|0\rangle\langle \chi|\right),
$$
where $\rho_{\chi 0}$ is given by (\ref{rhochi0rho0chi}). Non-zero contribution to interaction with the laser here gives the second term, equal to
\begin{equation}\label{stage3}
\rho_3=is{\rm Re}\left(\frac{1}{\mu_{20}}\right)\frac{\gamma^{-}_{{\rm re,em}}-\gamma^{+}_{{\rm re,em}}}
{\gamma^{+}_{{\rm re,em}}+\gamma^{-}_{{\rm re,em}}-2s^2{\rm Re}\left(\frac{1}{\mu_{20}}\right)}\left(1-|\langle\psi,\chi\rangle|^2\right)\left(|\chi\rangle\langle 0|-|0\rangle\langle \chi|\right).
\end{equation}

In the regime under consideration ($s\ne0$) this term is non-zero. This term is largest when the laser amplitude is chosen as follows
$$
s^2=-\frac{\gamma^{+}_{{\rm re,em}}+\gamma^{-}_{{\rm re,em}}}{2{\rm Re}\left(\frac{1}{\mu_{20}}\right)}=\frac{1}{2}\left(\gamma^{-2}_{{\rm re,em}}+\gamma^{+2}_{{\rm re,em}}+\gamma^{-2}_{{\rm im,em}}+\gamma^{+2}_{{\rm im,em}}\right),
$$
In this case
$$
i{\rm Im}\,\rho_{\chi 0}=-\frac{i}{2\sqrt{2}}\frac{\gamma^{-}_{{\rm re,em}}-\gamma^{+}_{{\rm re,em}}}
{\sqrt{\gamma^{-2}_{{\rm re,em}}+\gamma^{+2}_{{\rm re,em}}+\gamma^{-2}_{{\rm im,em}}+\gamma^{+2}_{{\rm im,em}}}}.
$$

If the state $\rho_1$ was excited by purely coherent light when $N_{\rm em}(k)=0$, hence
$$
\gamma^{+}_{{\rm re,em}}=\gamma^{+}_{{\rm im,em}}=0,
$$
$$
\gamma^{-}_{\rm re, em}=\pi\int
|g_{\rm em}(k)|^2\delta(\omega_{\rm em}(k)-\varepsilon_2+\varepsilon_0)dk,
$$
$$
\gamma^{-}_{\rm im, em}=-\int
|g_{\rm em}(k)|^2\,{\rm P.P.}\,{1\over\omega_{\rm em}(k)-\varepsilon_2+\varepsilon_0}dk,
$$
then ${\rm Im}\,\rho_{\chi 0}$ is maximal. If we ignore the shift $\gamma^{-}_{\rm im, em}$ we get the simple expression
$$
\rho_{\chi 0}=-\frac{i}{2\sqrt{2}}.
$$

\medskip

\noindent{\bf Summary.}\quad We have shown that the model of light-harvesting complex as a degenerate system with absorption and interaction with photons and phonons describes excitation of dark states which will have long lifetime and will be visible in spectroscopic experiments.

Earlier it was shown \cite{tmf2014} that  degeneracy can be used for quantum amplification of exciton transport (the supertransport effect). Results of the present paper show that in this case, as a side effect of the supertransport, we will obtain coherent dark states with long lifetime. This result can be discussed in relation of phenomenon of quantum photosynthesis \cite{Engel}, \cite{SFOG} and experimental observation of dark states in photosynthetic systems \cite{Novoderezhkin}.

\medskip

\noindent{\bf Remark.}\quad  We have discussed manipulations with quantum states, in particular with dark states: excitation of the states by coherent fields and using of Lindblad dissipation for projections of the states to some subspaces of matrices; since the bright vectors for different fields ($\chi$ and $\psi$ in the present model) can be non-parallel the manipulations under consideration are nontrivial. The considered manipulations using degeneracy and different ways of interaction with the degenerate energy level of the system are of very special form.

Manipulations of quantum states can be used for quantum computations \cite{OhyaVolovich}.
Long lifetime of dark states could help to avoid decoherence in quantum computers. For another proposal to reduce decoherence in quantum computers by manipulating with macroscopic parameters see \cite{Vol-2}. Different ideas of how to use photosynthetic quantum effects for computations are discussed in \cite{Engel}, \cite{Mohseni}, \cite{Kauffman}. 

It would be interesting also to discuss a possible connection of
manipulations with quantum states considered in this our work with those studied  in quantum control theory, see for example \cite{Pechen}.

\medskip

\noindent{\bf Quantum beats.}\quad Let us recall the model of interaction of two level system with laser. Dynamics of the density matrix of a system with levels $|0\rangle$, $|\chi\rangle$  (where $\|\chi\|=1$) is described by equation
$$
\frac{d}{dt}\rho(t)=i[\rho(t), H],\qquad H=s\sigma_x=s\pmatrix{ 0&  1 \cr 1 & 0 \cr},\qquad s\in\mathbb{R}.
$$

The basis of sigma--matrices is
$$
\sigma_0=|\chi\rangle\langle \chi|+|0\rangle\langle 0|,\qquad
\sigma_x=|\chi\rangle\langle 0|+|0\rangle\langle \chi|,\qquad
\sigma_y=i\left(-|\chi\rangle\langle 0| + |0\rangle\langle \chi|\right),\qquad
\sigma_z=|\chi\rangle\langle \chi|-|0\rangle\langle 0|.
$$
Operators $\sigma_0$, $\sigma_x$ are invariant with respect to the dynamics. Action of the generator of the dynamics in the subspace
$$
\rho=r(|\chi\rangle\langle \chi| -|0\rangle\langle 0|) + u i\left(-|\chi\rangle\langle 0| + |0\rangle\langle \chi|\right)
$$
is given by
$$
i[\cdot, H]\pmatrix{r \cr u}=
\pmatrix{ 0&  2s \cr
         -2s & 0 \cr}
\pmatrix{r \cr u}.
$$

The corresponding dynamics reduces to oscillations in this two-dimensional space with Rabi frequency $\Omega=2s$.

\bigskip

\noindent{\bf Acknowledgments.}\quad This research was funded by a grant from the Russian Science Foundation (Project No. 14-50-00005).

\end{document}